\documentclass[article, twocolumn, preprintnumbers, amsmath, amssymb,  superscriptaddress]{revtex4}

\usepackage[dvipsnames]{xcolor}
\usepackage{graphicx}
\usepackage{bbm}

\newcommand{\He}{$^{3} $He}

\begin{document}

\title{Polarization-Analyzed Small-Angle Neutron Scattering with an \textit{in-situ} $ ^{3} $He neutron spin filter at the China Spallation Neutron Source}

\author{Long Tian}
\thanks{These authors contributed equally to this work.}
\affiliation{Institute of High Energy Physics, Chinese Academy of Science, Beijing 100049, China}
\affiliation{Spallation Neutron Source Science Center, Dongguan 523803, China}
\affiliation{Guangdong Provincial Key Laboratory of Extreme Conditions, Dongguan 523803, China}

\author{Han Gao}
\thanks{These authors contributed equally to this work.}
\affiliation{Institute of High Energy Physics, Chinese Academy of Science, Beijing 100049, China}
\affiliation{Spallation Neutron Source Science Center, Dongguan 523803, China}
\affiliation{Guangdong Provincial Key Laboratory of Extreme Conditions, Dongguan 523803, China}
\affiliation{Center for Neutron Scattering and Advanced Light Sources, Dongguan University of Technology, Dongguan, Guangdong 523808, China}

\author{Tianhao Wang}
\thanks{These authors contributed equally to this work.}
\affiliation{Institute of High Energy Physics, Chinese Academy of Science, Beijing 100049, China}
\affiliation{Spallation Neutron Source Science Center, Dongguan 523803, China}
\affiliation{Guangdong Provincial Key Laboratory of Extreme Conditions, Dongguan 523803, China}

\author{Haiyun Teng}
\affiliation{Institute of High Energy Physics, Chinese Academy of Science, Beijing 100049, China}
\affiliation{Spallation Neutron Source Science Center, Dongguan 523803, China}

\author{Jian Tang}
\affiliation{Institute of High Energy Physics, Chinese Academy of Science, Beijing 100049, China}
\affiliation{Spallation Neutron Source Science Center, Dongguan 523803, China}
\affiliation{University of the Chinese Academy of Sciences, Beijing 100049, China}

\author{Qingbo Zheng}
\affiliation{Institute of High Energy Physics, Chinese Academy of Science, Beijing 100049, China}
\affiliation{Spallation Neutron Source Science Center, Dongguan 523803, China}
\affiliation{University of the Chinese Academy of Sciences, Beijing 100049, China}

\author{Taisen Zuo}
\affiliation{Institute of High Energy Physics, Chinese Academy of Science, Beijing 100049, China}
\affiliation{Spallation Neutron Source Science Center, Dongguan 523803, China}

\author{Tengfei Cui}
\affiliation{Spallation Neutron Source Science Center, Dongguan 523803, China}
\affiliation{Graduate School of China Academy of Engineering Physics, Beijing 100193, China}

\author{Bin Wang}
\affiliation{Guangdong Provincial Key Laboratory of Extreme Conditions, Dongguan 523803, China}
\affiliation{Center for Neutron Science and Technology, Guangdong Provincial Key Laboratory of Magnetoelectric Physics and Devices, School of Physics, Sun Yat-Sen University, Guangzhou, Guangdong 510275, China}

\author{Xu Qin}
\affiliation{Spallation Neutron Source Science Center, Dongguan 523803, China}
\affiliation{Center for Neutron Science and Technology, Guangdong Provincial Key Laboratory of Magnetoelectric Physics and Devices, School of Physics, Sun Yat-Sen University, Guangzhou, Guangdong 510275, China}

\author{Yongxiang Qiu}
\affiliation{Institute of High Energy Physics, Chinese Academy of Science, Beijing 100049, China}
\affiliation{Spallation Neutron Source Science Center, Dongguan 523803, China}

\author{Yuchen Dong}
\affiliation{Institute of High Energy Physics, Chinese Academy of Science, Beijing 100049, China}
\affiliation{Spallation Neutron Source Science Center, Dongguan 523803, China}
\affiliation{University of the Chinese Academy of Sciences, Beijing 100049, China}

\author{Yujie Zheng}
\affiliation{Institute of High Energy Physics, Chinese Academy of Science, Beijing 100049, China}
\affiliation{Spallation Neutron Source Science Center, Dongguan 523803, China}
\affiliation{Guangdong Provincial Key Laboratory of Extreme Conditions, Dongguan 523803, China}

\author{Zecong Qin}
\affiliation{Institute of High Energy Physics, Chinese Academy of Science, Beijing 100049, China}
\affiliation{Spallation Neutron Source Science Center, Dongguan 523803, China}
\affiliation{Guangdong Provincial Key Laboratory of Extreme Conditions, Dongguan 523803, China}

\author{Zehua Han}
\affiliation{Institute of High Energy Physics, Chinese Academy of Science, Beijing 100049, China}
\affiliation{Spallation Neutron Source Science Center, Dongguan 523803, China}

\author{Junpei Zhang}
\email{zhangjunpei@ihep.ac.cn}
\affiliation{Institute of High Energy Physics, Chinese Academy of Science, Beijing 100049, China}
\affiliation{Spallation Neutron Source Science Center, Dongguan 523803, China}
\affiliation{Guangdong Provincial Key Laboratory of Extreme Conditions, Dongguan 523803, China}

\author{He Cheng}
\email{chenghe@ihep.ac.cn}
\affiliation{Institute of High Energy Physics, Chinese Academy of Science, Beijing 100049, China}
\affiliation{Spallation Neutron Source Science Center, Dongguan 523803, China}

\author{Xin Tong}
\email{tongxin@ihep.ac.cn}
\affiliation{Institute of High Energy Physics, Chinese Academy of Science, Beijing 100049, China}
\affiliation{Spallation Neutron Source Science Center, Dongguan 523803, China}
\affiliation{Guangdong Provincial Key Laboratory of Extreme Conditions, Dongguan 523803, China}

\begin{abstract}
	Polarization-analyzed small-angle neutron scattering (PASANS) is an advanced technique that enables the selective investigation of magnetic scattering phenomena in magnetic materials and distinguishes coherent scattering obscured by incoherent backgrounds, making it particularly valuable for cutting-edge research. The successful implementation of PASANS in China was achieved for the first time at the newly commissioned Very Small Angle Neutron Scattering (VSANS) instrument at the China Spallation Neutron Source (CSNS). This technique employs a combination of a double-V cavity supermirror polarizer and a radio frequency (RF) neutron spin flipper to manipulate the polarization of the incident neutrons. The scattered neutron polarization is stably analyzed by a specially designed \textit{in-situ} optical pumping \He\ neutron spin filter, which covers a spatially symmetric scattering angle coverage of about 4.8$^{\circ}$. A comprehensive PASANS data reduction method, aimed at pulsed neutron beams, has been established and validated with a silver behenate powder sample, indicating a maximum momentum transfer coverage of approximately 0.25 \AA$^{-1}$.
\end{abstract}

\maketitle

\section{Introduction}
As a method of studying material structure on nano- to micro-scales, small-angle neutron scattering (SANS) complements X-ray and light scattering techniques, as neutrons exhibit high transmittance, sensitivity to magnetism and nuclei, and is commonly used in material research, such as proteins, polymers and magnetic nanoparticles\cite{Guo1990,Baker2009,Milde2013,McCulloch2013,Jonietz2010}. When neutron spins are polarized, due to the different interactions between polarized neutrons and materials, SANS is further allowed to distinguish the contributions of nuclear coherent scattering, incoherent scattering and magnetic scattering\cite{Moon1969,Sch1993,Squires1978}, which benefits for the measurements requiring to extract accurate structural parameters or clarfiy the magnetic domain distributions in materials\cite{Krycka2013,Krycka2010,Das2022,Gaspar2010}. Polarized neutron small angle scattering instruments are therefore required to be applied at the neutron facilities worldwide.

With the development of the \He\ neutron spin filter (NSF)\cite{Qin2021,Huang2021,Parnell2009,Coulter1990,Jiang2013}, the challenge of analyzing the polarization of scattered neutron beams by the polarization-analyzed small-angle neutron scattering (PASANS) method was overcome. The properties of a \He\ NSF can be customized for a specific neutron wavelength band by optimizing the geometrical parameters and \He\ gas pressure in the \He\ NSF, enabling the polarization of neutrons for both pulsed or single wavelength neutron beam to be analyzed after scattering by a sample. Furthermore, the \He\ NSF offers better accessibility for scattered neutrons in a SANS instrument compared to the polarizing supermirror array by increasing the coverage of the analyzing angle. This is a crucial aspect of SANS experiments, as it affects the range of reciprocal space that can be measured and available scale of samples when applying the PASANS.

The \He\ NSF has primarily been used on neutron beamlines in \textit{ex-situ} mode, such as the Very Small Angle Neutron Scattering (VSANS) diffractometer at the National Institute of Standards and Technology (NIST) Center\cite{Chen2023}, the KWS-1 instrument at the Heinz Maier-Leibnitz Zentrum\cite{Feoktystov2015}, and the D33 instrument at the Institut Laue-Langevin\cite{Dewhurst2016}. The \textit{ex-situ} \He\ NSF is shielded from spatial stray magnetic fields by a cylindrical $\mu$-metal casing. Its compact size makes it feasible for deployment in SANS experiments, as it occupies little space between the sample and the detector array. However, powerful neutron sources, such as reactor sources, and longer polarization decay lifetimes of the NSF are required due to the intrinsic decay of the \He\ polarization in \textit{ex-situ} \He\ NSF. Additionally, continuous calibration of the \He\ polarization prior to sample measurements, or periodic calibration during ongoing measurements, is necessary to ensure the accuracy of corrections for each dataset when switching to a new NSF, which inevitably increases the complexity of data reduction. As a complement to the \textit{ex-situ} mode, the \textit{in-situ} \He\ NSF allows one to maintain a constant polarization of \He\ through the use of \textit{in-situ} optical pumping, thereby greatly simplifying the measurement procedure, and making PASANS an available technique at pulsed neutron sources\cite{Zhang2022,Jiang2014,Hayashida2017,Okudaira2020,Salhi2019}.

In this paper, we report the design and the first successful application of the PASANS technique in China by utilizing polarized neutron instruments, including a customized \textit{in-situ} NSF at the Very Small Angle Neutron Scattering (VSANS) instrument at the China Spallation Neutron Source (CSNS). Additionally, a comprehensive method was established for the correction of polarized neutron data for PASANS measurements with a pulsed neutron beam, which has been validated through measurements using silver behenate powders.

\section{Experimental Instruments}
\subsection{PASANS setup on the VSANS beamline}
The VSANS at CSNS has set its research goal in the physical design to cover the studies of materials that requires the polarized neutron scattering technique. As shown in Fig. 1, a rotational exchange drum is mounted upstream of the VSANS beamline, which enabling the precise exchange of three channels: neutron guide, flight tube, and polarizing supermirror. This design enables switching between polarized and unpolarized modes. The polarizing supermirror is specifically designed to polarize cold neutrons in the range of 2.2 \AA\ to 11 \AA\ by employing a double-V cavity composed of \textit{m} = 5 Fe/Si supermirrors, optimized based on simulation results 
\begin{figure}[h]
	\centering
	\includegraphics[scale=0.65]{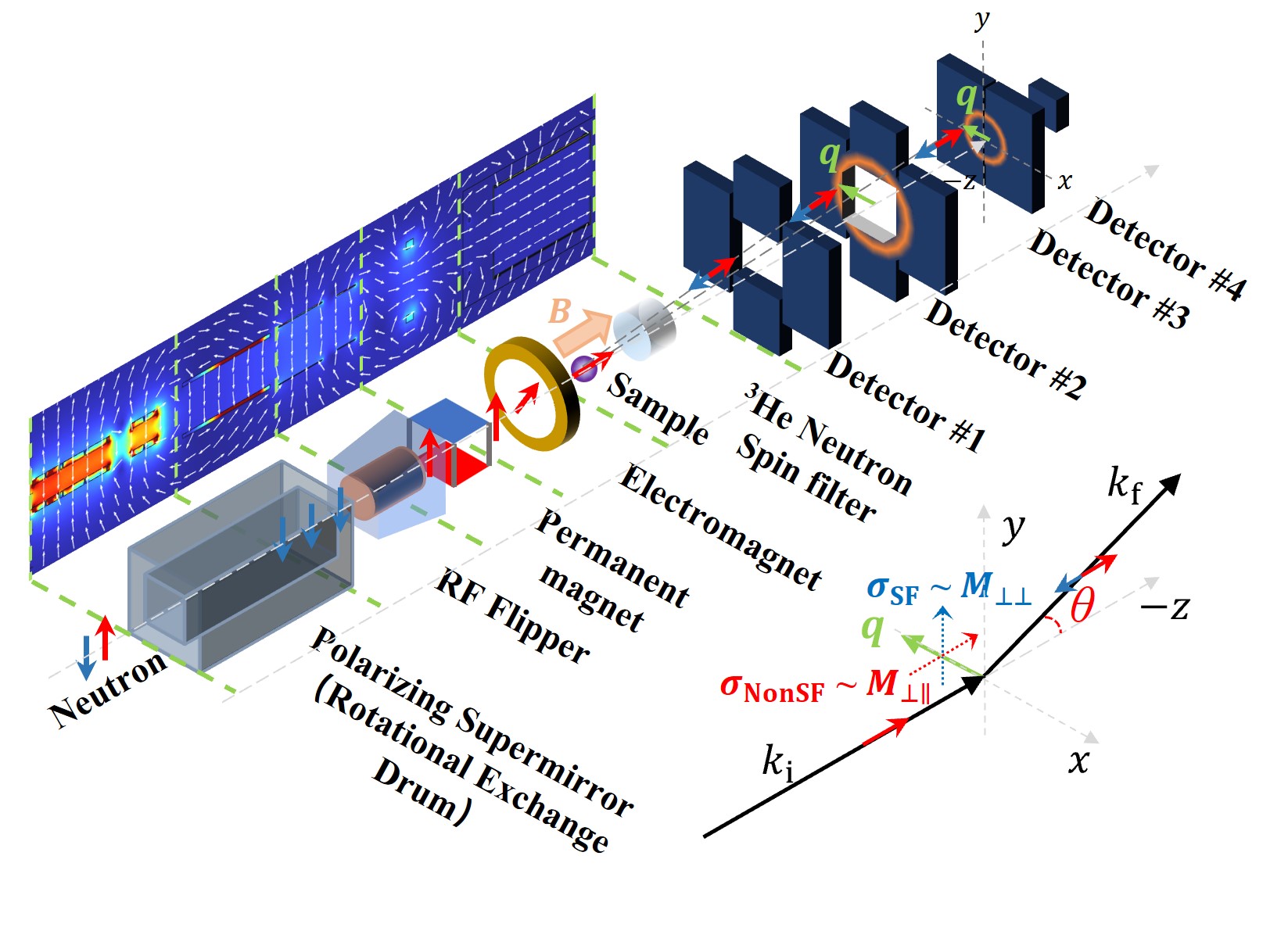}
	\caption{Diagram of the PASANS setup utilized at the VSANS beamline at CSNS. The blue and red arrows represent the neutron polarization along the neutron path ($k_{\mathrm{i}}$ is the incident neutron wave vector aligned along -z axis, $k_{\mathrm{f}}$ is the scattered neutron wave vector), permanent magnets and electromagnets are applied to generate polarized neutron guide fields to maintain and manipulate the polarization direction. The green arrow represents the momentum transfer \textit{q} during the neutron scattering with the sample. Red and blue solid arrows indicate the polarization of incident and scattered neutrons, and the dashed arrows refer to the projected magnetic moment components ($M_{\perp \parallel}$ or $M_{\perp \perp}$) that contribute to the spin-flip ($\sigma_{\mathrm{SF}}$) or non-spin-flip ($\sigma_{\mathrm{NonSF}}$) ~{scattering}. The red $\theta$ indicates the neutron scattering angle. }
	\label{fig:example1}
\end{figure}
\noindent
for polarizing efficiency and transmission ratio \cite{Zuo2024}. The vertical neutron polarization guide field, which is used to maintain neutron polarization after the supermirror, is generated by a combination of permanent magnet bars and iron plate yokes mounted on the downstream flight tubes. Furthermore, a radio frequency (RF) neutron spin flipper manufactured by SwissNeutronics Inc. is utilized to flip the polarization of incident neutrons by a $ \pi $ angle. 

The static magnetic field environment beside the flipper, as shown in Fig. 1, is optimized to create a compatible gradient field distribution through the finite element method to meet the magnetic resonance condition. Simulation results indicate a flipping efficiency of 98\% at 2.2 \AA, which simplifies the polarized neutron data reduction process (explained in the following section). An \textit{in-situ} \He\ NSF system is specially designed for VSANS as the analyzer, which is essential for PASANS since the neutron absorption cross-section of \He\ is spin-dependent and the \He\ NSF can be tailored to have a larger neutron scattering angle coverage for the scattered beam. Additionally, the neutron polarization is manipulated adiabatically by customized guide fields along the beam path\cite{Tian2023}, indicated by the colored arrows in Fig. 1, rotated by 90° before the analyzer. The \textit{in-situ} pumping method for maintaining the \He\ NSF on the beamline ensures stable and continuous analyzing capability throughout the measurements. Photographed in Fig. 2(a), the PASANS experimental setup at VSANS was developed based on the design described above.

\subsection{\textit{In-situ} $ ^{3} $He neutron spin filter}
The \textit{in-situ} \He\ NSF at CSNS was first developed and delivered to neutron beamlines in 2021\cite{Zhang2022}, and it has been successfully utilized as the neutron spin polarizer and analyzer for neutron imaging and reflectometry measurements at CSNS\cite{Salman2022}. To accommodate the geometry of the conically scattered neutron beam for the SANS instrument, a new generation of \textit{in-situ} \He\ NSF, termed as \textit{in-situ} - SANS, was developed based on our innovative prototype designs. The optical-pumping cell (OPC) features an inner diameter of 72 mm and a length of 80 mm, as shown in Fig. 2(b), facilitating a scattering angle of approximately 4.8$^{\circ}$ (corresponds to $\theta$ in the yellow cone) and a maximum \textit{q} value of about 0.24 \AA$^{-1}$ when the cell is positioned 37 cm from the sample. Notably, the dimensions of the optics, oven windows, magnetic shielding cavity, and box on their exit sides have been enlarged to allow the analyzed scattered neutron beam to pass through the NSF setup unimpededly. These significant changes in geometry pose challenges in improving and maintaining the \He\ polarization. Additionally, neutron-absorbing materials were attached to the exit side of the NSF box to minimize scattering interference beyond the cell coverage area. The figure of merit (FOM) of the OPC produced for PASANS measurements is optimized to be 11.01 bar$\cdot$cm to match the beam profile of VSANS, which has a neutron wavelength band of 2.2 \AA\ to 6.7 \AA. We employ the Free Induction Decay (FID) method to track the evolution of \He\ polarization, validating a saturated polarization level of 62.3\% with the Electron Paramagnetic Resonance (EPR) method, corresponding to a \He\ polarization of 61.1\% calibrated through neutron transmission measurements (see Fig. 3(a) and (b)). As shown in Fig. 2(a), the new \textit{in-situ} - SANS also allows for flexible experimental configurations. The main \He\ NSF box can be moved in or out of the beam path by operating the support cart's motor\cite{Tang2025}, enabling a switch between half-polarized SANS, PA-
\begin{figure}[h]
	\centering
	\includegraphics[scale=0.75]{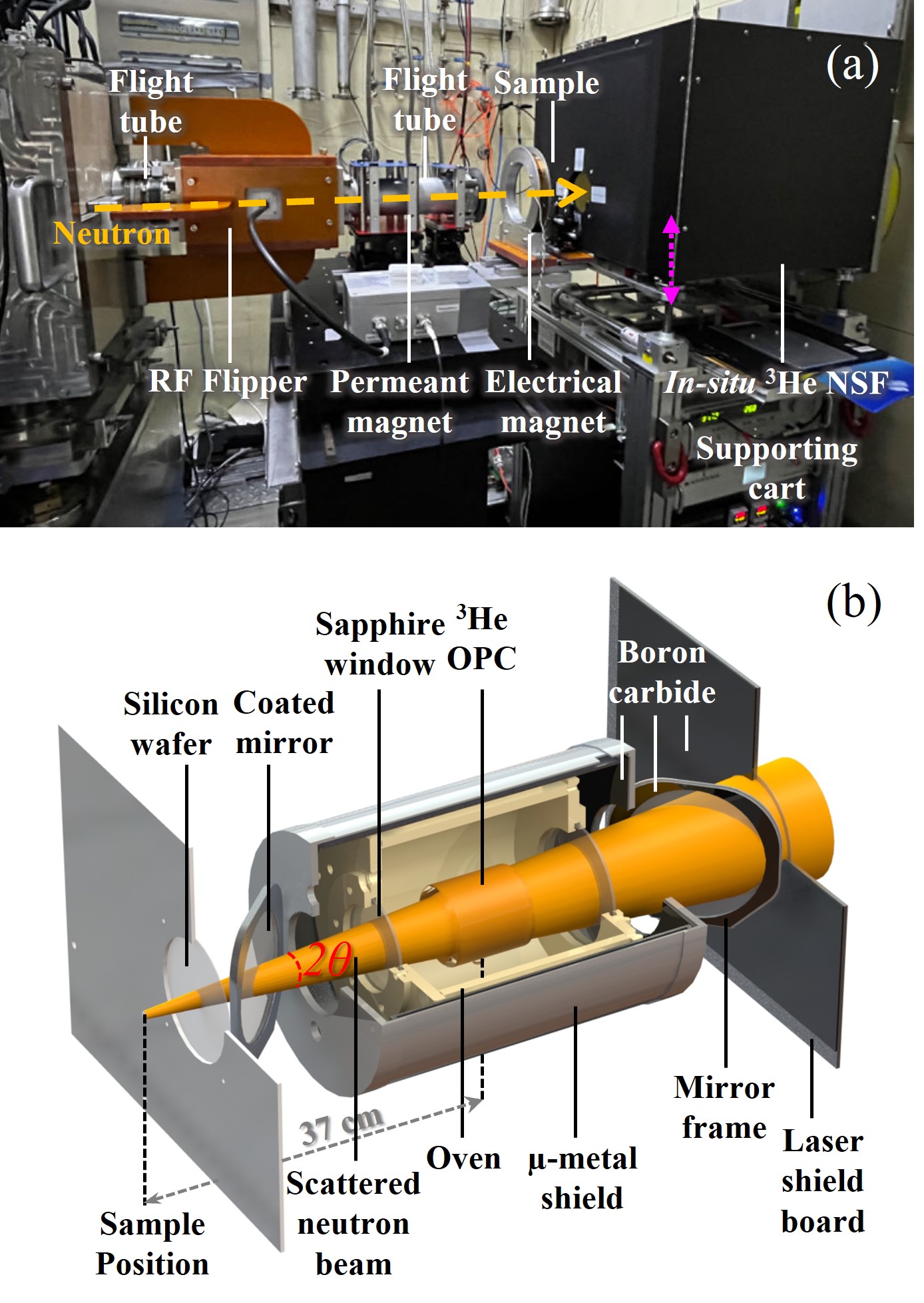}
	\caption{(a) Photograph of the polarized neutron setup utilized on VSANS beamline at CSNS. The polarizing supermirror is mounted upstream of the sample position at the first drum of the beamline, which is not shown here. The main part of \textit{in-situ} - SANS, which is shown as the enclosed black box, can be adjusted continuously along the magenta dashed arrow to achieve different experimental modes. (b) The cross section of the \textit{in-situ} - SANS. The scattered neutron beam, represented by a yellow cone, passes through the windows and mirrors (made of silicon single crystal or sapphire) and the \He\ cell, and the red $\theta$ denotes twice the neutron scattering angle ($\theta$ ~ 4.8$^{\circ}$) shown in Fig.1. Boron carbide is coated at the exit to shield the scattered neutrons that exceed the angular coverage of \He\ cell. The dashed gray line with arrows indicates the distance between the sample and the \He\ OPC.} 
	\label{fig:example1}
\end{figure}
\noindent
SANS, and unpolarized SANS. These features underscore the exceptional suitability of \textit{in-situ} - SANS for PASANS applications.

\section{Polarized Data Reduction}
In neutron scattering theory, the total scattering amplitude comprises distinct contributions \cite{Blume1963, Maleyev1963}: (1) the nuclear coherent and isotope incoherent scattering, represented by the term \textit{N}, which are independent of neutron polarization; (2) the nuclear spin-incoherent scattering \textit{I}, which is dependent on neutron polarization and has an approximate probability of 2:1 for flipping or not flipping the neutron spins; and (3) the magnetic scattering term \textit{M$_{\perp}$}, which depends on both neutron spin and the scattering vector, and can be further separated into \textit{M$_{\perp \parallel}$} and \textit{M$_{\perp \perp}$}, where the subscripts indicate that the sample magnetization component is perpendicular to the scattering vector (\textit{q}) and can be either parallel or perpendicular to the neutron polarization, depending on whether the neutron spin flips.
To summarize, concerning the neutron spin dependence, the total scattering can be divided into non-spin-flip (NonSF) and spin-flip (SF) scattering components, represented by the red and blue arrows in Fig. 1, respectively. The NonSF scattering comprises the terms \textit{N}, 1/3$\cdot I$, and \textit{M$_{\perp \parallel}$}, while the SF scattering consists of 2/3$\cdot I$, and\textit{ M$_{\perp \perp}$}.

In a typical polarized neutron scattering process, information about the NonSF and SF scattering in a sample is collected by manipulating the polarization of the incident and scattered neutrons. Additionally, the polarized neutron data is necessary to be corrected for unintended scattering leakage from incorrect spin states in the raw data due to the inefficiencies of polarized neutron instruments. To extract the absolute sample scattering amplitude, a derivation based on the efficiencies of polarized neutron instruments is required \cite{Krycka2012, Nambu2023, Wildes1999}. For SANS measurements conducted at a pulsed neutron source, the scattering intensity \textit{S}(x, y, $\lambda$) is a function of the two-dimensional position within the scattering plane (x, y) and the neutron wavelength $\lambda$, which will be simplified as \textit{S} hereafter. NonSF scattering $S _{++} $, $S _{--} $, and SF scattering $S _{+-} $, $S _{-+} $, where the subscript indicates the spin of the incident or scattered neutron as being parallel ($ + $) or antiparallel ($ - $) to the guide field direction, in PASANS are typically given as follows:
\begin{align}
	S_{++} & = \zeta_{\mathrm{sm +}}^{+} \zeta_{\mathrm{^3He +}}^{+} \sigma_{+ +} + \zeta_{\mathrm{sm +}}^{-} \zeta_{\mathrm{^3He +}}^{+} \sigma_{- +} \nonumber \\
	& + \zeta_{\mathrm{sm +}}^{-} \zeta_{\mathrm{^3He +}}^{-} \sigma_{- -} + \zeta_{\mathrm{sm +}}^{+} \zeta_{\mathrm{^3He +}}^{-} \sigma_{+ -} \\
	S_{--} & = \zeta_{\mathrm{smf -}}^{+} \zeta_{\mathrm{^3He -}}^{+} \sigma_{+ +} + \zeta_{\mathrm{smf -}}^{-} \zeta_{\mathrm{^3He -}}^{+} \sigma_{- +} \nonumber \\
	& + \zeta_{\mathrm{smf -}}^{-} \zeta_{\mathrm{^3He -}}^{-} \sigma_{- -} + \zeta_{\mathrm{smf -}}^{+} \zeta_{\mathrm{^3He -}}^{-} \sigma_{+ -}
\end{align}
and 
\begin{align}
	S_{+-} & = \zeta_{\mathrm{sm +}}^{+} \zeta_{\mathrm{^3He -}}^{+} \sigma_{+ +} +  \zeta_{\mathrm{sm +}}^{-} \zeta_{\mathrm{^3He -}}^{+} \sigma_{- +} \nonumber \\
	& + \zeta_{\mathrm{sm +}}^{-} \zeta_{\mathrm{^3He -}}^{-} \sigma_{- -} + \zeta_{\mathrm{sm +}}^{+} \zeta_{\mathrm{^3He -}}^{-} \sigma_{+ -} \\
	S_{-+} & = \zeta_{\mathrm{smf -}}^{+} \zeta_{\mathrm{^3He +}}^{+} \sigma_{+ +} + \zeta_{\mathrm{smf -}}^{-} \zeta_{\mathrm{^3He +}}^{+} \sigma_{- +} \nonumber \\
	& + \zeta_{\mathrm{smf -}}^{-} \zeta_{\mathrm{^3He +}}^{-} \sigma_{- -} + \zeta_{\mathrm{smf -}}^{+} \zeta_{\mathrm{^3He +}}^{-} \sigma_{+ -} 
\end{align}
where the $\sigma_{\pm \pm}$ refers to the simplified description of sample scattering cross sections $\sigma_{\pm \pm}(x,y,\lambda)$ with different neutron spin states, $\zeta_{**}^{*}$ represents the probability that a spin-up (+) or spin-down (-) neutron can pass through a polarized neutron instrument. $\zeta_{**}^{*}$ can also be expressed in terms of the combination of the wavelength-dependent instrument polarization parameters \textit{$ P_{*} $}($ \lambda $) and the transmission ratio \textit{$ T_{*} $}($ \lambda $), simplified as
\begin{align}
	& \zeta_{\mathrm{sm/^3He +}}^{+} = (\frac{1+P_{\mathrm{sm/cell}}}{2})T_{\mathrm{smf/^3HePol}}\\
	& \zeta_{\mathrm{sm/^3He +}}^{-} = (\frac{1-P_{\mathrm{sm/cell}}}{2})T_{\mathrm{smf/^3HePol}}\\
	& \zeta_{\mathrm{smf/^3He -}}^{+} = (\frac{1-P_{\mathrm{smf/cell}}}{2})T_{\mathrm{smf/^3HePol}}\\
	& \zeta_{\mathrm{smf/^3He -}}^{-} = (\frac{1+P_{\mathrm{smf/cell}}}{2})T_{\mathrm{smf/^3HePol}}
\end{align}
where $ P_{\mathrm{smf}} $ = $ P_{\mathrm{sm}}P_{\mathrm{f}} $, $ T_{\mathrm{smf}} $ = $ T_{\mathrm{sm}}T_{\mathrm{f}} $. The $ P_{\mathrm{sm/cell/f}} $ refers to the polarizing or analying efficiency of a supermirror ($ P_{\mathrm{sm}}$), $ ^{3} $He spin filter ($ P_{\mathrm{cell}}$), and the flipping efficiency of the neutron spin flipper ($ P_{\mathrm{f}}$). The $ T_{\mathrm{sm/^3HePol/f}} $ refers to the transmission ratio of an unpolarized neutron beam passing through the polarized neutron devices. However, both $T_{\mathrm{sm}}$ and $T_{\mathrm{f}}$ are reduced in our case since the experimental data ($ S_{**} $) are normalized by the prime beam measured with the supermirror in place. In addition, $T_{\mathrm{f}}$ = 1 as a RF flipper is utilized. 

According to the above equations, to extract the scattering cross section of samples, the instrument parameters \textit{$ P_{*} $} and \textit{$ T_{*} $} need to be calibrated in advance by conducting direct transmission measurements with unpolarized neutron beam. Benefiting from the time-independent $ ^{3} $He polarization of an \textit{in-situ} $ ^{3} $He NSF, the poalrizing efficiencies of the instruments can be denoted as
\begin{align}
	& P_{\mathrm{cell}} = \sqrt{1-(\frac{T_{\mathrm{^3HeDepol}}}{T_{\mathrm{^3HePol}}})^{2}}\\
	& P_{\mathrm{sm}} = \frac{I_{++}-I_{+-}}{P_{\mathrm{^3He}}}\\
	& P_{\mathrm{f}} = \frac{I_{--}-I_{-+}}{I_{++}-I_{+-}}
\end{align}
where the $ T_{\mathrm{^3HePol}} $ and $ T_{\mathrm{^3HeDepol}} $ represent the transmissions of an unpolarized neutron beam passing through a polarized or depolarized $ ^{3} $He cell, and $ I_{**} $ refers to the neutron transmission ratio for different neutron spin states. All the \textit{$ P_{*} $} and \textit{$ T_{*} $} will be calibrated before each experimental cycle on the beamline.

Furthermore, measurements of block-beam ($ S^{\mathrm{bk}} $, spin-independent), the transmission ratio of the sample ($ T_{\mathrm{s}} $) and the scattering from the sample holder ($ S_{**}^{\mathrm{h}} $) are also necessary to determine the absolute scattering intensity of the sample. Since a typical sample holder is always non-magnetic, its spin-independent scattering measurements could hereby be reduced when the $ P_{\mathrm{f}} $ is assumed as 1, because of its $ \sigma_{++}^{\mathrm{h}} = \sigma_{--}^{\mathrm{h}}$, $ \sigma_{+-}^{\mathrm{h}} = \sigma_{-+}^{\mathrm{h}}$. Additionally, both $ S_{**}^{\mathrm{h}} $ and $ S_{**}^{\mathrm{s}} $ need to be normalized to remove the effects of incident neutron flux and $T_{\mathrm{sm}}$ before the polarization correction. The absolute scattering intensity of the sample in reciprocal space can then be expressed as
\begin{align}
	\sigma_{++}^{\mathrm{s}}(q) = \sum_{x,y,\lambda \rightarrow q}\frac{B_{1} S_{++}^{\mathrm{cor}}+B_{4} S_{+-}^{\mathrm{cor}}+A_{3} S_{-+}^{\mathrm{cor}}+A_{2} S_{--}^{\mathrm{cor}}}{C}\notag \\
	\sigma_{-+}^{\mathrm{s}}(q) = \sum_{x,y,\lambda \rightarrow q}\frac{B_{3} S_{++}^{\mathrm{cor}}+B_{2} S_{+-}^{\mathrm{cor}}+A_{1} S_{-+}^{\mathrm{cor}}+A_{4} S_{--}^{\mathrm{cor}}}{C}\notag \\
	\sigma_{--}^{\mathrm{s}}(q) = \sum_{x,y,\lambda \rightarrow q}\frac{B_{2} S_{++}^{\mathrm{cor}}+B_{3} S_{+-}^{\mathrm{cor}}+A_{4} S_{-+}^{\mathrm{cor}}+A_{1} S_{--}^{\mathrm{cor}}}{C}\notag \\
	\sigma_{+-}^{\mathrm{s}}(q) = \sum_{x,y,\lambda \rightarrow q}\frac{B_{4} S_{++}^{\mathrm{cor}}+B_{1} S_{+-}^{\mathrm{cor}}+A_{2} S_{-+}^{\mathrm{cor}}+A_{3} S_{--}^{\mathrm{cor}}}{C}\label{eq:second}
\end{align} 
where
\begin{align}
	& S_{**}^{\mathrm{cor}} = S_{**}^{\mathrm{s}}/T_{\mathrm{sh}}-T_{\mathrm{s}} S_{**}^{\mathrm{h}}/T_{\mathrm{h}}\notag\\
	& A_{1} = (P_{\mathrm{cell}}+1)(P_{\mathrm{sm}}+1)\notag\\
	& A_{2} = (P_{\mathrm{cell}}-1)(P_{\mathrm{sm}}-1)\notag\\
	& A_{3} = (P_{\mathrm{cell}}+1)(P_{\mathrm{sm}}-1)\notag\\
	& A_{4} = (P_{\mathrm{cell}}-1)(P_{\mathrm{sm}}+1) \notag\\
	& C = 2P_{\mathrm{cell}}P_{\mathrm{sm}}(P_{\mathrm{f}}+1)T_{\mathrm{^3HePol}}\label{eq:second}
\end{align}
$ T_{\mathrm{sh}} $ and $ T_{\mathrm{h}} $ refer to the transmission ratios of the sample with holder and an empty holder, respectively. $ B_{1} $ to $ B_{4} $ are presented by replacing the $ P_{\mathrm{sm}} $ with $ P_{\mathrm{smf}} $ in $ A_{1} $ to $ A_{4} $ in equation (13). For non-magnetic materials, such as polymers and proteins, we have \textit{M$_{\perp \parallel}$} = \textit{M$_{\perp \perp}$} = 0, which further simplifies the equation (12) to $\sigma_{++}^{\mathrm{s}} = \sigma_{--}^{\mathrm{s}}$ and $\sigma_{+-}^{\mathrm{s}} = \sigma_{+-}^{\mathrm{s}}$.

\section{Results and Discussion}
\subsection{Performance of the PASANS setup}
The polarized neutron efficiencies of the instruments were calibrated through neutron transmission measurements without samples on the VSANS at CSNS. Neutron data from PASANS measurements were collected with a wavelength range from 2.2 \AA\ to 6.7 \AA, employing a collimation length of 8.31 m. The $ \# $3 detector array was used to collect the transmission data by positioning it 12.2 m from the sample stage, while a 2 mm pinhole B$ _{4} $C slit was mounted before the sample position to define the beam size. Figure 3(a) shows the wavelength dependence of $ T_{\mathrm{^3HePol}} $ and $ P_{\mathrm{cell}} $, where the latter is derived from equation (9). To verify the stabilization of the polarization analyzing capability of the \textit{in-situ} - SANS, we measured $T _{\mathrm{^3HePol}} $ approximately every 30 hours. A difference of less than 0.4\% in $T _{\mathrm{^3HePol}} $ was observed, which corresponds to a fluctuation of $ P_{\mathrm{^3He}} $ of less than 1.2\%, and the weighted 
\begin{figure}[h]
	\centering
	\includegraphics[scale=0.65]{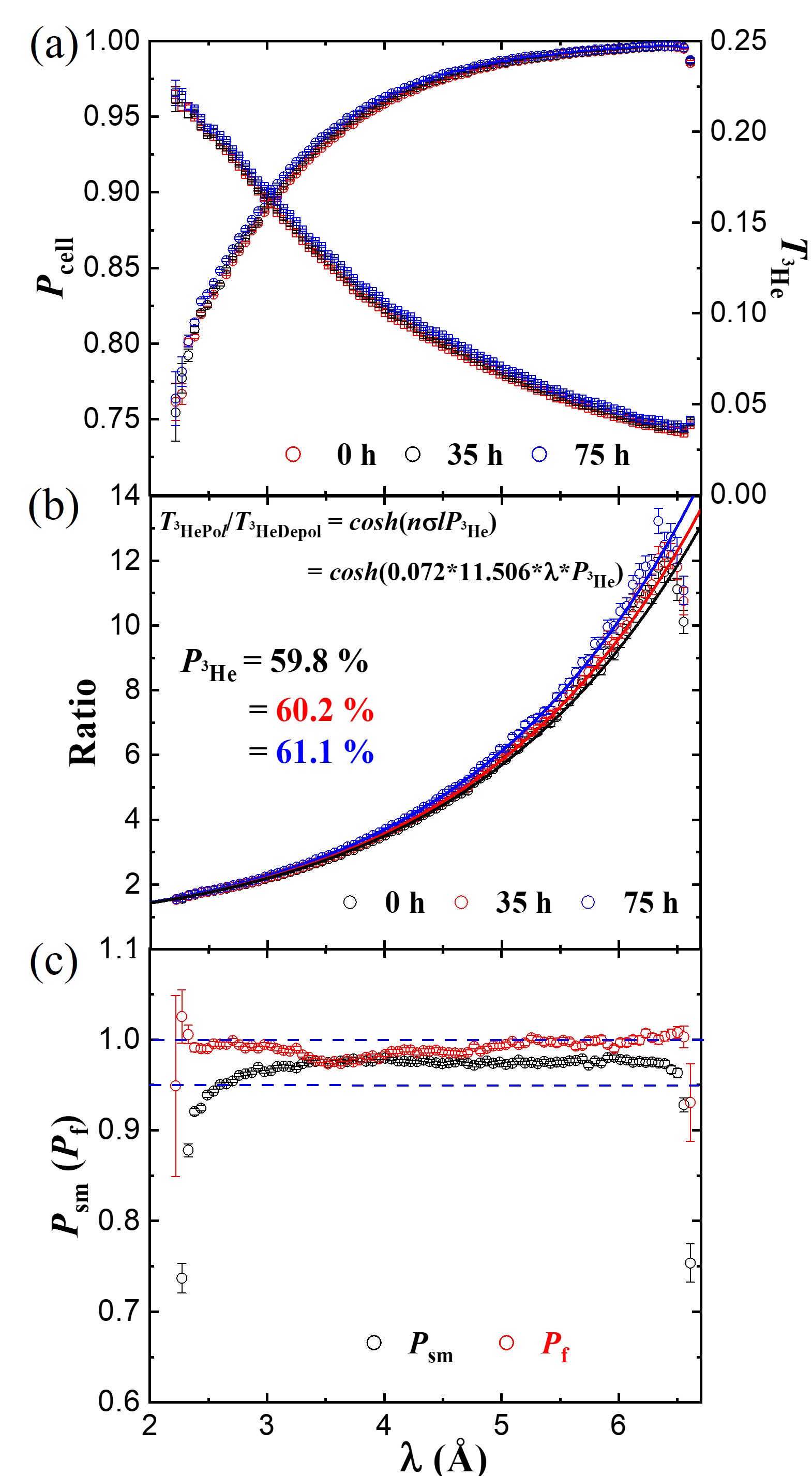}
	\caption{Neutron wavelength dependence of polarized neutron efficiencies of the devices. (a) The neutron polarization analyzing ability and polarized neutron transmission ratio of the \textit{in-situ} \He\ NSF, measured at 0 h, 35 h and 75 h. (b) Wavelength dependence of the polarized neutron transmission ratio parameter $ T_{\mathrm{^3HePol}}/T_{\mathrm{^3HeDepol}} $. The solid lines represent the best-fit curves for determining the \He\ polarization $ P_{\mathrm{^3He}} $ using equations shown in the subfigure. (c) Wavelength dependence of the polarizing efficiency of the supermirror and the flipping efficiency of the RF flipper. The dashed red lines are the guide lines showing the boundaries of 95\% and 100\%. }
	\label{fig:example1}
\end{figure}
\noindent
average difference in $ P_{\mathrm{cell}} $ was approximately 0.5\% based on the prime beam flux distribution. 
This indicates that the \He\ NSF can be used as a stable analyzer at the beamline, allowing $ P_{\mathrm{cell}} $ and $T _{\mathrm{^3HePol}} $ to be treated as constants during the 
polarization correction process. The corresponding saturated 
\He\ polarization $ P_{\mathrm{^3He}} $ was also fitted to 61.1\% $\pm$ 0.1\% during the experiment \cite{Zhang2022,Huang2021} (see Fig. 3(b)).

The polarizing efficiency ($ P_{\mathrm{sm}} $) of the supermirror was calibrated according to equation (10) by flipping the \He\ polarization while keeping the flipper off. As shown in Fig. 3(c), $ P_{\mathrm{sm}} $ exceeds 95\% at 2.6 \AA\ and reaches about 97.5\% at longer wavelengths, which is similar to our simulation result of $ P_{\mathrm{sm}} > $ 95\% at 2.4 \AA, and the difference may arise from the collimation accuracy of the installation. The spin-dependent transmission ratio ($ T_{\mathrm{sm}} $) of the supermirror also exceeds 32\% across the experimental wavelength range. Moreover, the flipping efficiency ($ P_{\mathrm{f}} $) of the RF flipper mounted at the VSANS beamline was optimized to be over 98\% for neutron wavelengths above 2.2 \AA\ by adjusting the RF power. The excellent performance of the flipper allows us to simplify the polarized neutron data reduction process during the experiment.

\subsection{Sample measurement}

Silver behenate (AgBE) has been well established as a standard 
sample for wavelength calibration in SANS instruments \cite{Gilles1998}, as its first three Bragg reflection peaks are accessible within the SANS scattering angle range. Since it is nonmag-
\begin{figure}[h]
	\centering
	\includegraphics[scale=0.7]{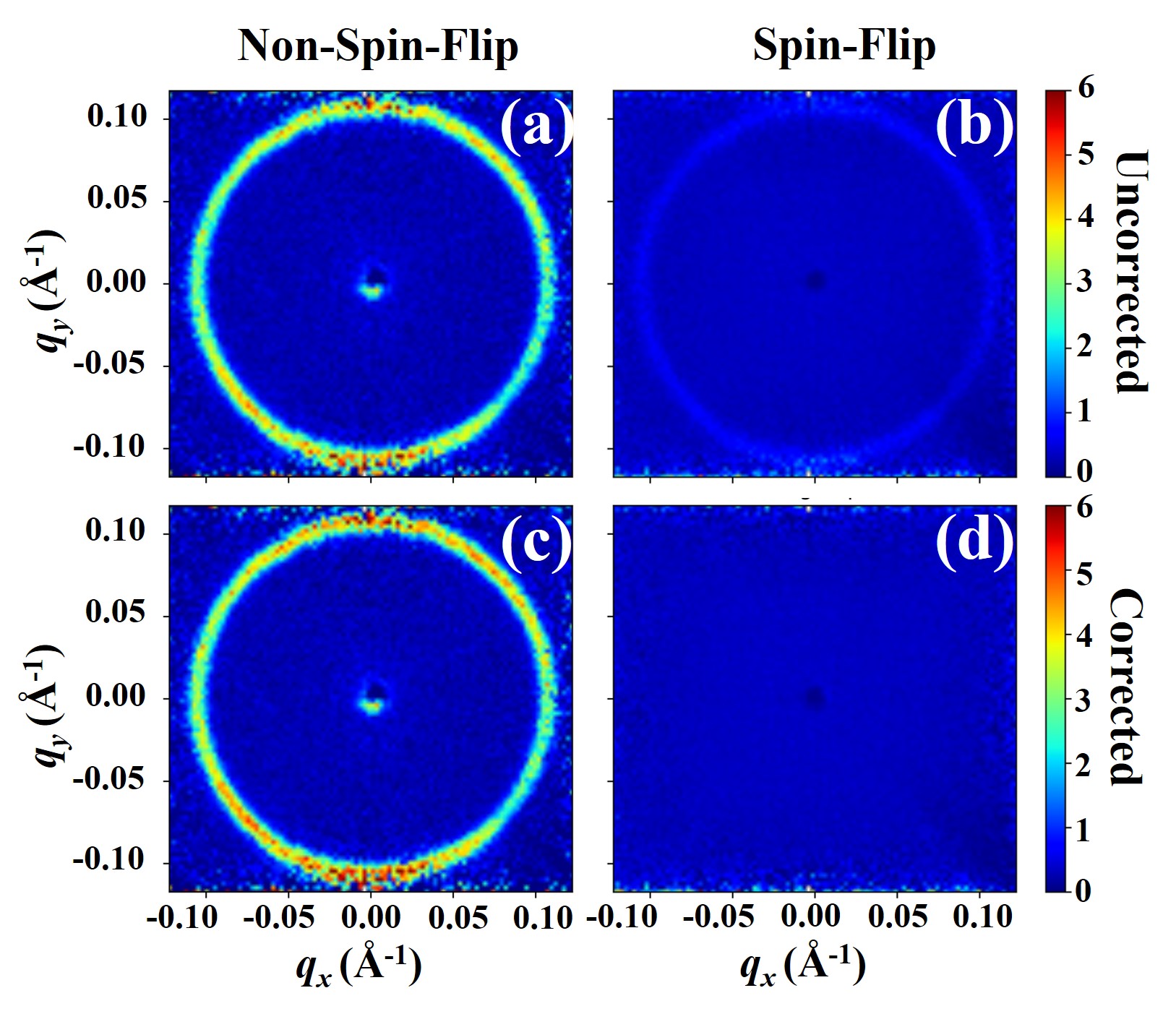}
	\caption{Diffraction patterns of AgBE powder collected by $ \# $3 detectors. (a) and (b): Non-spin-flip ($ \sigma_{--}^{\mathrm{s}}(q) $) and spin-flip ($ \sigma_{+-}^{\mathrm{s}}(q) $) scattering patterns after absolute intensity normalization, except the PASANS correction. (c) and (d): PASANS corrected scattering patterns. }
	\label{fig:example1}
\end{figure}
\noindent
netic and hydrogen-rich, its nuclear scattering components \textit{N} and \textit{I} can be effectively distinguished 
through NonSF and SF measurements. The PASANS experiment at VSANS was conducted with silver behenate powder sample encased in quartz cells with an optical path length of 2 mm, placed 4.5 m and 12 m away from the middle-angle ($ \# $2) and small-angle 
($ \# $3) detectors, respectively. Considering the OPC geometry, the polarization-analyzed scattered neutrons will be detected 
by the entire $ \# $3 detector and part of the $ \# $2 detector (Fig. 1).

Fig. 4 shows the comparison between the NonSF and SF results measured over a neutron wavelength range of 2.2 \AA\ to 6.7 \AA. A distinct isotropic ring at \textit{q} $\sim$ 0.11 \AA $ ^{{-1}} $ was observed in both two-dimensional images in the reciprocal space of the $ \# $3 and $ \# $2 detectors, corresponding to nuclear coherent scattering at small \textit{q}. Furthermore, the second Bragg peak at \textit{q} $\sim$ 0.22 \AA\ $ ^{{-1}} $ was captured by the inner part of the middle-angle detector. Compared to the NonSF, the SF also exhibits weak isotropic scattering at the same \textit{q} positions, induced by the leakage of scattered neutrons with incorrect spin states due to the non-100\% efficiency of the polarized neutron instruments.
\begin{figure}[h]
	\centering
	\includegraphics[scale=0.68]{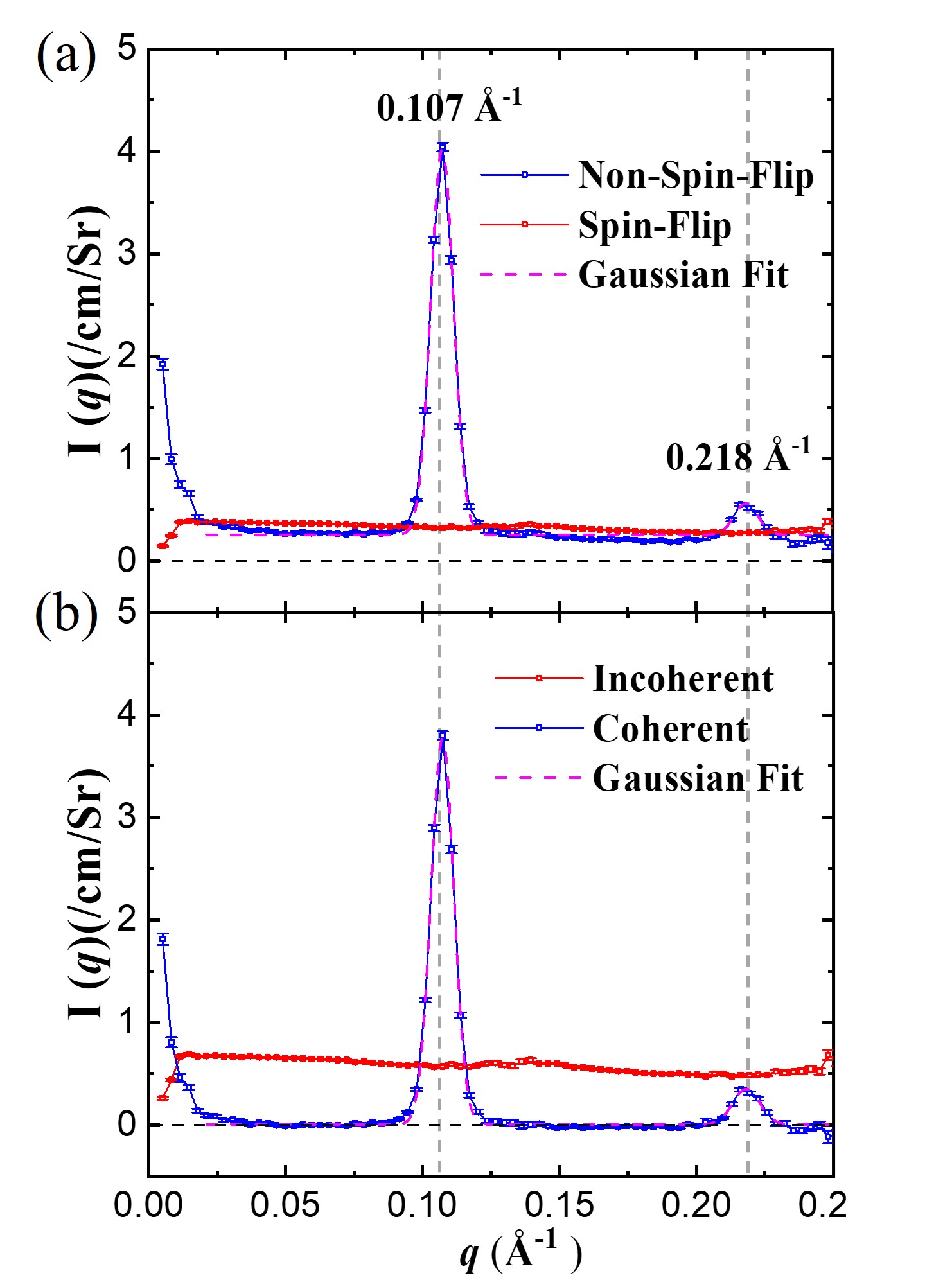}
	\caption{Radially averaged scattering data on an absolute scale after PASANS correction on both $ \# $2 and $ \# $3 detectors. Corrected spin-flip and non-spin-flip scattering curves are indicated as empty blue and red squares in (a). Separated nuclear spin incoherent and coherent scattering curves shown in red and blue empty squares in (b). Two diffraction peaks in the non-spin-flip scattering and nuclear coherent scattering curves are fitted using a Gaussian equation, depicted as a magenta dashed line. }
	\label{fig:example1}
\end{figure}
\noindent

By implementing the polarized data correction method introduced in the above section, the scattering intensity corresponding to each spin state is allocated to the correct scattering state. Fig. 4(c) and (d) illustrate the corrected NonSF and SF information in reciprocal space, where the nuclear coherent scattering in the SF has been removed, making SF scattering information homogeneous and independent of \textit{q}. The azimuthally averaged absolute intensity of the corrected NonSF and SF curves is displayed in Fig. 5(a). Two peaks stand out in the NonSF curve, dominated by nuclear spin-coherent scattering, in contrast to the flat SF curve, which is contributed solely by nuclear spin-incoherent scattering. To estimate the effect of multiple scattering in a hydrogen-rich thick sample, the contribution weight of \textit{I} in the SF curve is 
denoted as \textit{m}, which can be determined by fitting the ratio of $\sigma^{\mathrm{s}}_{- -}$ to $\sigma^{\mathrm{s}}_{- +}$ in the low \textit{q} range, where the \textit{N} contribution in the NonSF curve could be neglected. Our measurements yield a ratio of \textit{m} = 0.569, 
consistent with the strong multiple scattering occurring in a thicker sample. Fig. 5(b) also presents the comparison of calculated \textit{N} and \textit{I} based on equations (14-15):
\begin{align}
	\sigma_{++}^{\mathrm{s}}(\mathrm{q}) = \sigma_{--}^{\mathrm{s}}(\mathrm{q}) = N + (1-m)I \\
	\sigma_{-+}^{\mathrm{s}}(\mathrm{q}) = \sigma_{+-}^{\mathrm{s}}(\mathrm{q}) = mI
\end{align}
where the nuclear spin-coherent and spin-incohernet scattering are well separated in silver behenate below \textit{q}$\sim$0.25 \AA $ ^{{-1}} $.

\section{Conslusion}
In this paper, we report the first successful implementation of the PASANS technique at the newly commissioned VSANS at CSNS. The polarized neutron equipment has been deployed along the neutron path to accomplish the full process of neutron spin polarizing, flipping, and analyzing in PASANS. By utilizing the \textit{in-situ} \He\ NSF as the neutron spin analyzer, a large symmetric scattering cross-section coverage was achieved in both real and reciprocal spaces, with a maximum \textit{q} value of approximately 0.25 \AA $ ^{{-1}} $ under an incident neutron wavelength range of 2.2 \AA\ to 6.7 \AA . The efficiencies of the polarizing supermirror, spin flipper, and \textit{in-situ} \He\ NSF have been calibrated, demonstrating a highly efficient and stable neutron polarization capability on VSANS. Furthermore, a detailed process for polarized neutron data reduction is introduced, considering a pulsed neutron beam as the incident beam. Silver behenate powder was used as a standard sample during the PASANS commissioning, revealing distinct differences between its nuclear spin-coherent scattering and spin-incoherent scattering, proving the feasibility of PASANS in the neutron scattering research field in China.

Building on the interfaces developed initially for upgrading the \textit{in-situ} \He\ NSF, future enhancements will focus on integrating the PASANS method with complex sample environments, including multi-axis magnetic fields with strong field strength. This integration is expected to broaden scientific applications in magnetic materials with complex magnetic orders, such as magnetic skyrmions, by utilizing polarized neutrons on VSANS.
\\

\noindent\textit{Acknowledgements:}
This work was supported by the GuangDong Basic and Applied Basic Research Foundation Grant (No. 2021B1515140016, No. DG22311526), the National Natural Science Foundation of China (No. 12075265 and U2032219), the Guangdong Natural Science Funds for Distinguished Young Scholars, the National Key Research and Development Program of China (Grant No. 2020YFA0406000), the Guangdong Provincial Key Laboratory of Extreme Conditions (Grant No. 2023B1212010002). We thank the staff members of the Very Small Angle Neutron Scattering at the China Spallation Neutron Source (CSNS) (https://csns.cn/31113.02.CSNS.VSANS), for providing technical support and assistance in data collection and analysis.

\end{document}